\documentclass[aps,prl,reprint]{revtex4-1}
\usepackage{graphicx}
\usepackage{dcolumn}
\usepackage{bm}
\usepackage{amsmath}
\usepackage{amssymb}
\usepackage{xcolor}

\begin{document}
\title{Observation of ultrafast laser-plasma evolution by pump-probe reflectometry and Doppler spectrometry}
\author{Amitava Adak$^{\dagger}$, Prashant Kumar Singh, Amit D. Lad, Gourab Chatterjee, G. Ravindra Kumar}
\email{grk@tifr.res.in}
\email{$^{\dagger}$adak.amitava@gmail.com}
\affiliation{Tata Institute of Fundamental Research, Dr. Homi Bhabha Road, Colaba, Mumbai-400005, India}

\begin{abstract}
We demonstrate pump-probe techniques, namely the Doppler spectrometry and the reflectometry in detail, which directly capture the time-resolved ultrafast evolution of high intensity femtosecond laser-driven hot, dense plasma. These techniques are capable of capturing ultrafast plasma dynamics on time scales of sub 100 femtosecond. We have shown the dynamics of high intensity femtosecond laser-driven shock like disturbance into the plasma at densities more than $10^{22}$ cm$^{-3}$.  This can help understanding the physics related to shock ignition, supernova explosion and many other astrophysical scenarios and also can have implications in medicine and chemistry. Furthermore, we have investigated the ultrafast acoustic phenomena due to hydrodynamics inside an expanding hot, dense plasma in its transient phase by the correlated measurements of Doppler spectrometry and reflectometry.  
\end{abstract}
\maketitle
\section{Introduction}
Creating `micro-star' on table top by high intensity laser-matter interaction has been a fascinating area of research because of its importance in basic plasma physics\cite{Kruerbook}, laboratory astrophysics\cite{Remington1999}, laser-fusion\cite{Ross2013} and laser driven shock wave physics\cite{Drakebook}. Intense femtosecond laser has the capability of producing inertially confined high energy density plasma on a solid surface. The temporal dynamics of such extreme states are of great interest for decades due to its relevance in basic sciences as well as in various applications. Despite of having many direct and indirect measurements and computer simulation techniques, this area of research still remains exciting and rapidly being enriched and widened by new discoveries\cite{AmitavaPRL2015}. Various pump-probe techniques are generally employed to create and measure the ultrafast dynamics of high intensity short pulse laser-plasma interaction and its consequences. For example, at decently high laser-intensity of $10^{15}$ W/cm$^2$, the plasma was characterized by using an independent probe reflectivity and the dynamical change in the thermal conductivity of the plasma was measured\cite{sandhu2005}. Another important study by Liu \textit{et al.}\cite{LiuPRL1992}, at laser-intensities of $10^{16}$ W/cm$^2$ observed the competition between the ponderomotive and thermal forces in a short-scale-length laser plasma measuring the Doppler shift in the reflected probe. A pump-probe reflectometry and Doppler spectrometry reveals the generation of high frequency acoustic oscillations\cite{AmitavaPRL2015} in a dense plasma created by a 800 nm, 30 fs laser pulse at focused peak intensity of $\sim 5\times10^{16}$ W/cm$^2$. In an another work\cite{AmitavaPOP2014}  at higher laser intensities of $10^{18}$ W/cm$^2$ we have shown the dynamical propagation of a shock-like-disturbance towards highly dense plasma (near solid-density) by pump-probe spectrometry using an ultraviolet probe. Our recent work\cite{shock-control} using these experimental techniques, reveal the possibility of controlling the femotosecond laser-driven shock which can potential applications in medicine and chemistry\cite{Dlott1999,Reed2007,ReedPRL2006,Takayama2004}.
\begin{figure}[t]
\includegraphics[width=0.9\columnwidth]{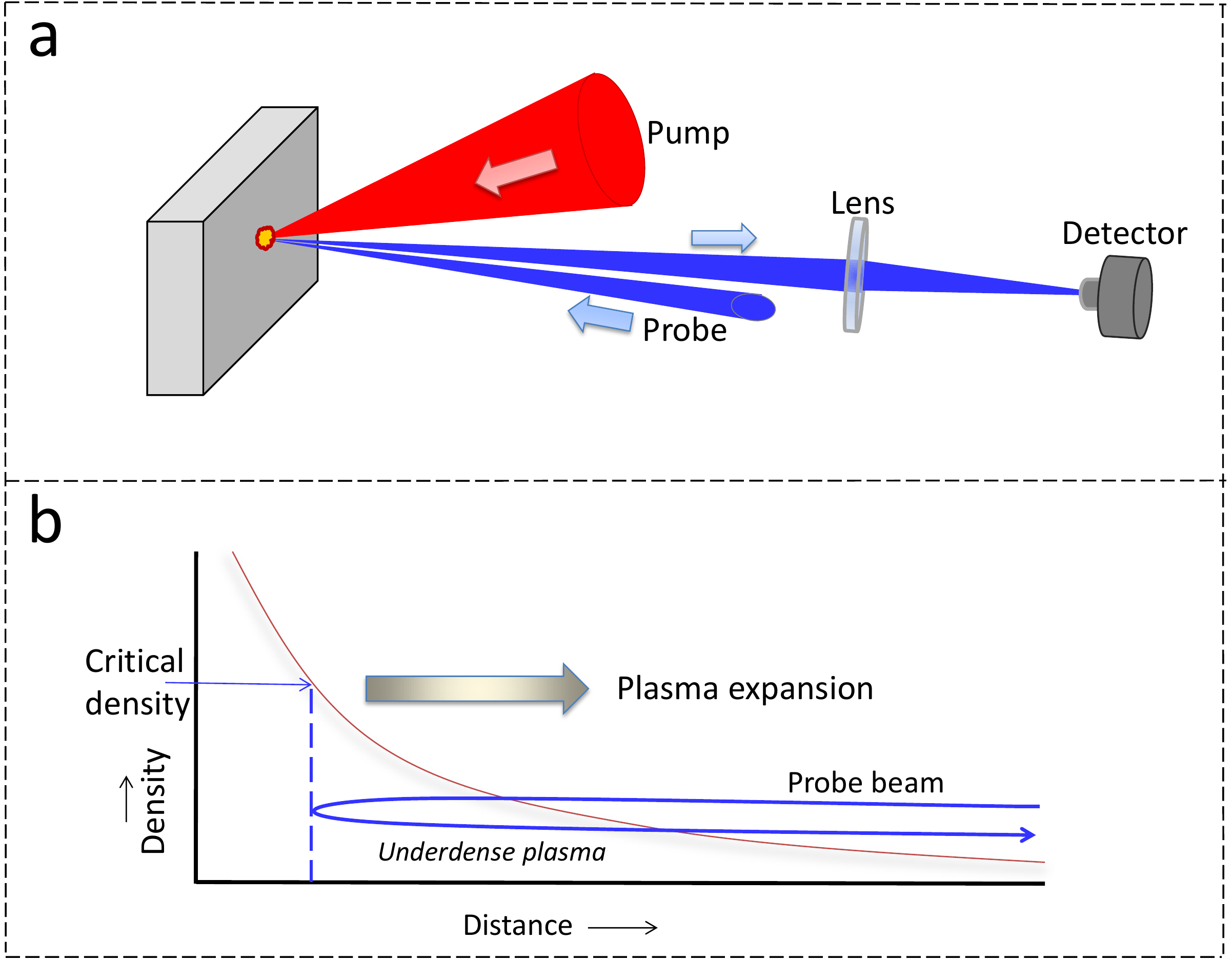}
\caption{a) Pump-probe setup for a femtosecond laser solid interaction. The pump laser (red) is focused on the solid slab. The spotsize was about 11 $\mu$m (FWHM). A probe (blue) was derived from the main pump pulse and focused on the interaction region after time delayed (in controlled manner) by a delay stage. The reflected probe was collected by the detectors like photodidode, high resolution spectrometers etc. b) A simplified picture of laser-solid interaction. The red curve is a typical density profile of an freely expanding plasma (towards vacuum, in this case right). The blue dashed line indicates the critical density layer for the normally incident probe (blue line). The probe travels through the underdense plasma and reflecs back from the corresponding critical density layer. The reflected probe experiences the dynamics near the critical density layer and the `changes' are measured by the detectors. }\label{setup}
\end{figure}

In this paper, we have demonstrated the techniques of pump-probe reflectometry and Doppler spectrometry in detail which helps revealing the ultrafast dynamics in a hot dense plasma created by an infrared ($\sim 800$ nm) short pulse (30 fs) laser at focused intensity in the non relativistic range $10^{16}$-$10^{17}$ W/cm$^2$ and the shock-dynamics at near relativistic intensities\cite{AmitavaPOP2014}. These simple pump-probe techniques mentioned above is now firmly established and helps doing fundamental physics such as finding new mechanism of ultrafast acoustic generation in hot, dense plasma\cite{AmitavaPRL2015}.

\section{Pump-probe setup}

\begin{figure}[]
\begin{center}
\includegraphics[width=\columnwidth]{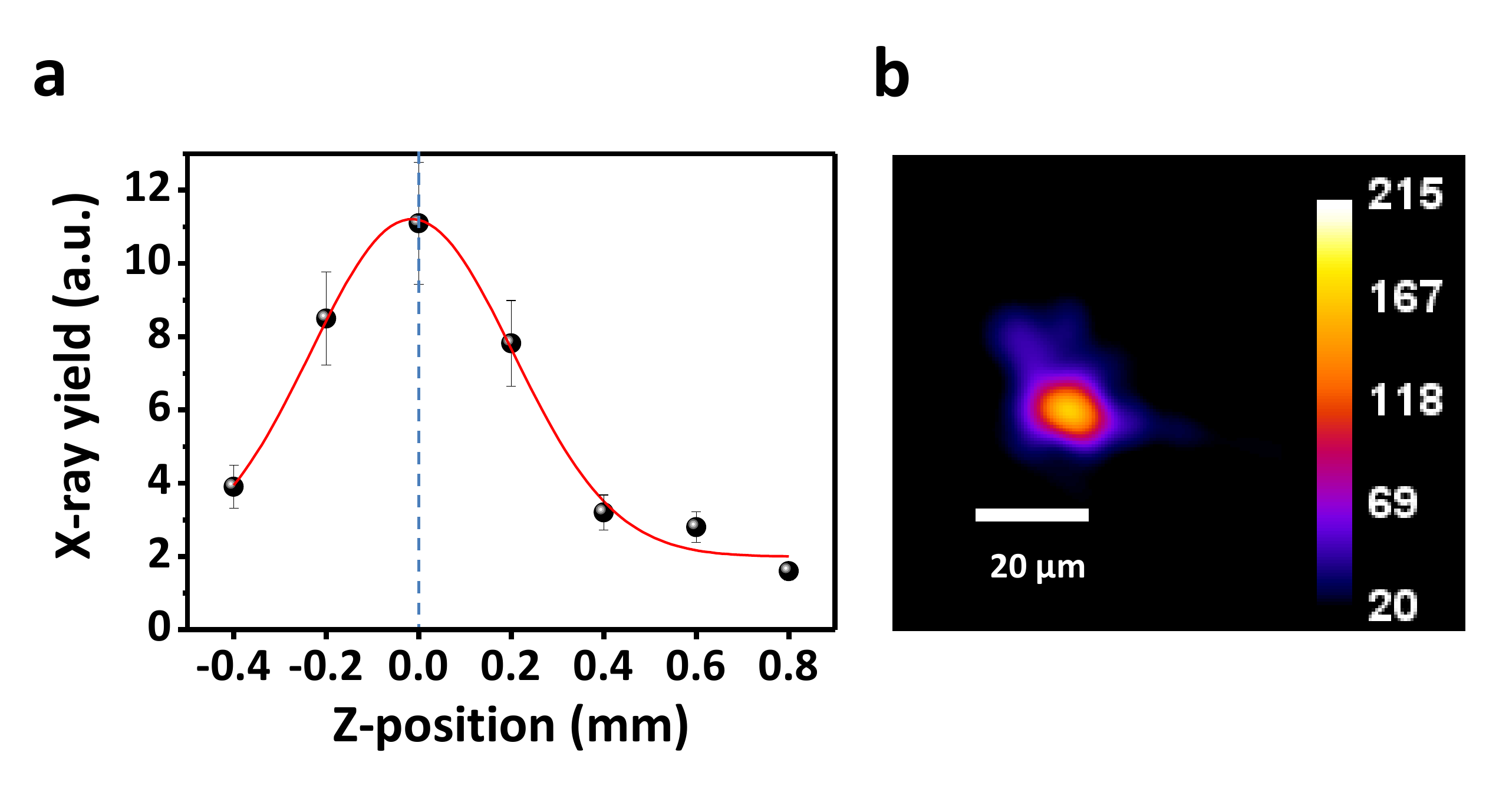}
\caption{ \textbf{(a)} Hard x-ray yield at various target positions. The dashed line indicates the position of the target producing the maximum hard x-ray yield, which is the final position of the target for the experiment. \textbf{(b)} The image of the focal spot measured by direct imaging of the focused beam on a CCD.} \label{ch2-z-focus-spot}
\end{center}
\end{figure} 
A typical pump-probe setup is shown in fig. \ref{setup}a. The pump pulse ($\sim$ 800 nm, 30 fs) was focussed on a plane solid target (BK7 glass) by an off-axis parabolic mirror (OAP) at $45^\circ$ angle of incidence in a f/4 optical geometry. We used a frequency resolved optical gating device (GRENOUILLE, Swamp Optics \& Newport Corp.) to measure the pulse duration of the short pulse laser in single shot mode and to characterize of the picosecond contrast of the laser we used a third-order cross-correlator (SEQUOIA, Amplitude Technologies). A probe pulse was derived from the same pump pulse, passed through a delay stage (to control the pump-probe temporal delay) and then upconverted to its second harmonic before focussing it to the interaction region on the target at near normal incidence. The focal spot of the probe pulse on the target was 50 $\mu$m (FWHM). The reflected probe was collected by another lens and fed into a photodectector to measure the reflectivity or into a high resolution spectrometer to measure its spectrum and hence the Doppler shift.

The experimental chamber was kept at a pressure of $\sim10^{-5}$ torr. The geometric position of the focal region (best focus) of the pump beam was determined by irradiating a target (Al coated BK-7 glass) kept at different distances from the OAP and looking at the hard x-ray signal (fig. \ref{ch2-z-focus-spot}a) generated by the interaction. The x-ray measurement was done at a laser intensity of $\sim 10^{16}$ W/cm$^2$ and the signal was collected by a NaI(Tl) scintillator placed outside the experimental chamber at a distance of 50 cm from the interaction point. Finally the target mounted on a xyz$\theta$ stage (Newport Corp.) was kept at the position corresponding to the maximum x-ray signal (indicated by arrow in fig. \ref{ch2-z-focus-spot}a). The intensity distribution of the focal spot was measured independently by attenuating the pump intensity and directly imaging on a charged couple device (CCD). 35\% of the total energy was found inside the full width at half maxima (11 $\mu$m) and about 50\% inside 15 $\mu$m. Figure \ref{ch2-z-focus-spot}b shows the intensity profile of the laser spot at best focus position, where different colors indicate different intensities.

\section{Pump-probe reflectometry}

\begin{figure}[b]
\includegraphics[width=\columnwidth]{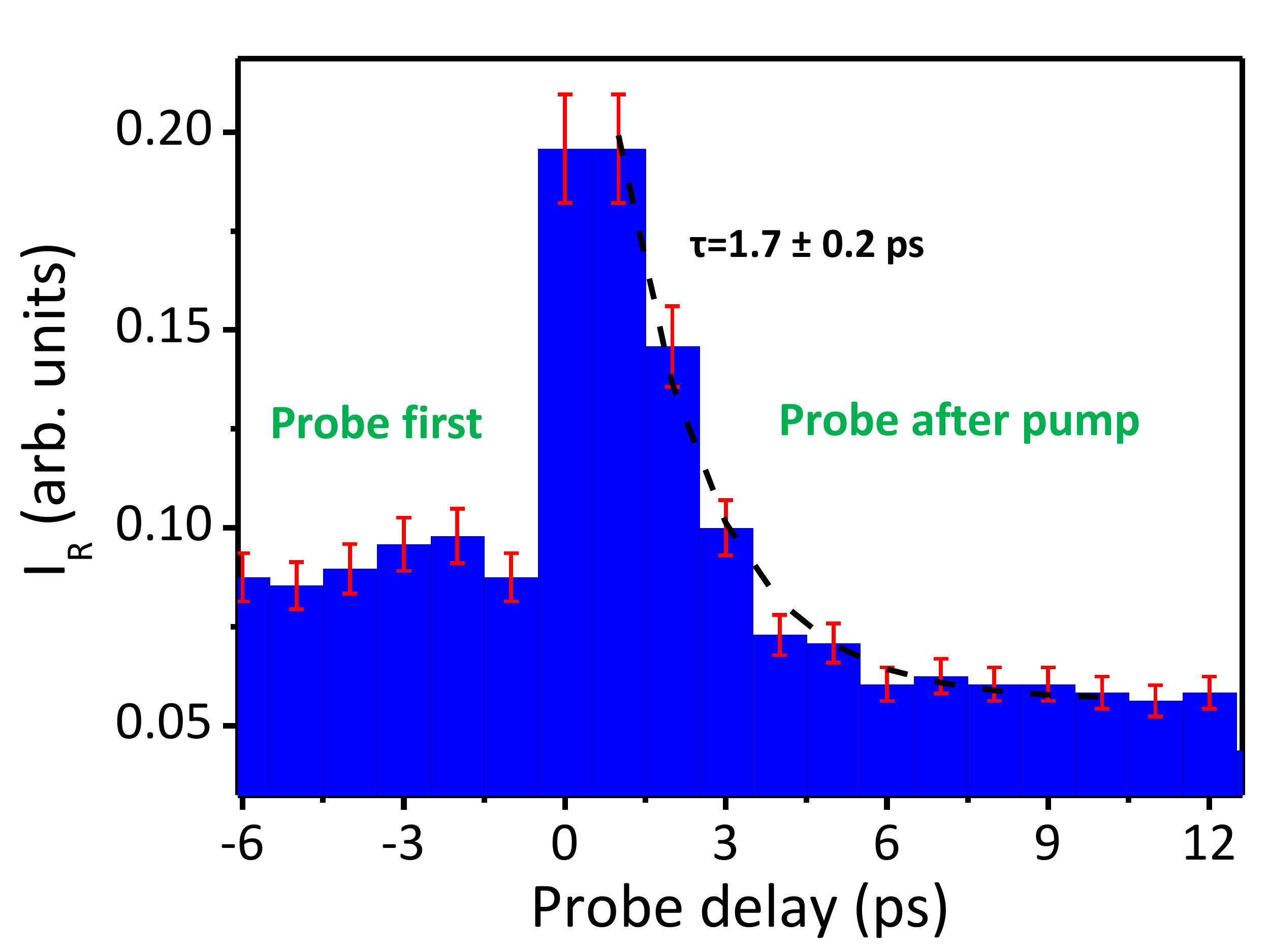}
\caption{Reflected probe intensity is plotted against the probe delay. The spike in reflectivity indicates the arrival of the pump pulse. The exponential decay of the reflectivity after zero delay is due to the probe absorption in the expanding plasma.}\label{rflspike}
\end{figure}

\begin{figure}
\includegraphics[width=\columnwidth]{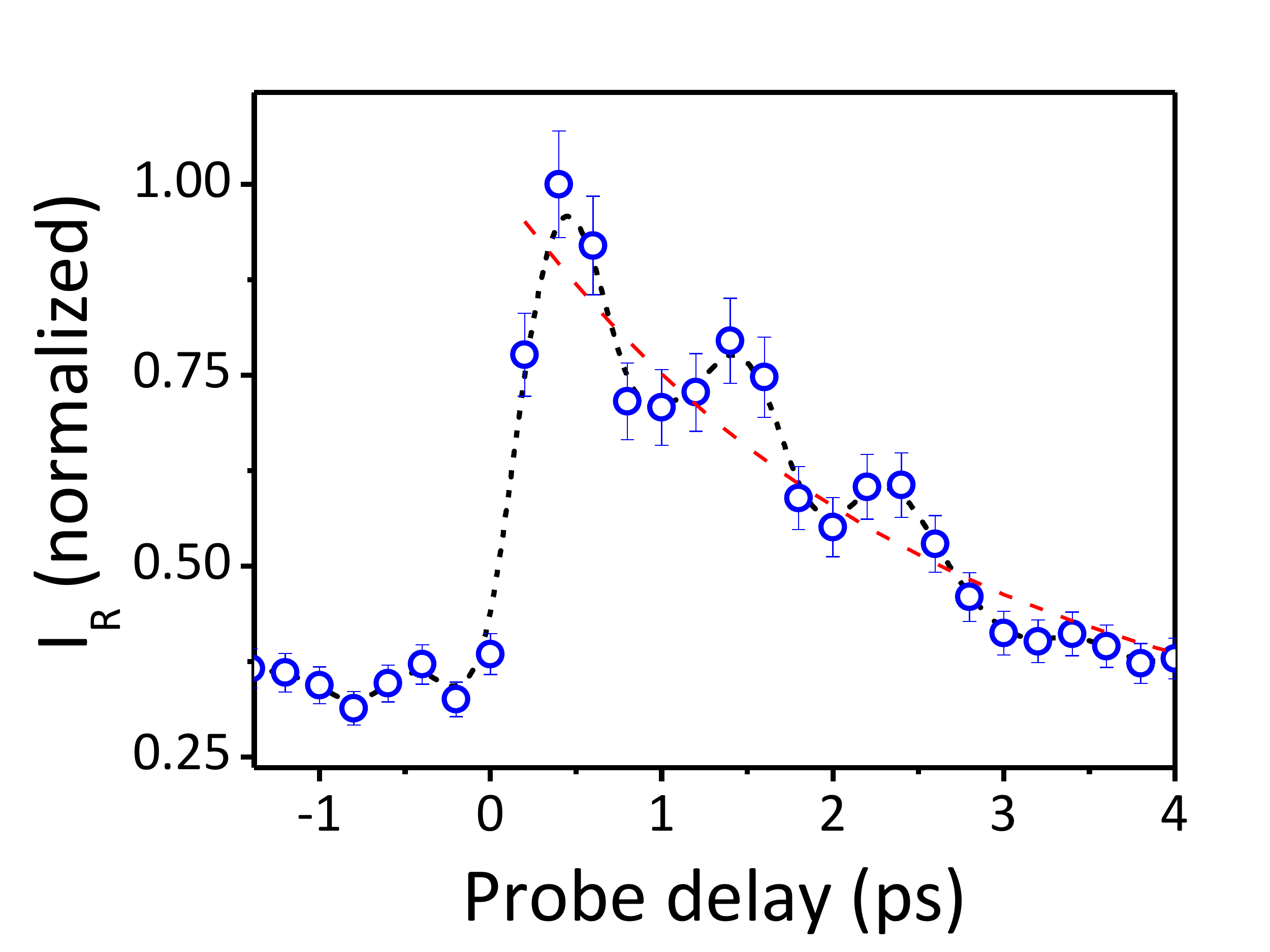}
\caption{Time resolved reflectivity of the second harmonic probe from the plasma shows oscillations in THz frequency. The plasma was produced on plane BK-7 glass surface at focused laser intensity of $9\times10^{16}$ W/cm$^2$.}\label{rfl-osc-2w}
\end{figure}

Plasma reflectivity measurements have been carried out for years to characterize many facets of intense laser-plasma interactions\cite{MilchbergPRL1988,NgPRL1994,WangOL1992,Benuzzi-MounaixPRE1999,sandhu2005,LiuPRL1992}. Sandhu \textit{et al.,}\cite{sandhu2005} studied the electrical resistivity in metal by observing the evolution of the probe reflectivity for 10s of ps and showed the resistivity saturation effect in the intermediate region of hot plasma and cold solid. Observation of ultrafast transport of laser-driven hot electrons on a dielectric surface is another important example of spatio-temporal pump-probe reflectivity measurement\cite{PrashantPOP2013} performed in our laboratory. Most of these pump-probe reflectivity measurements rely on the probe absorption in the expanding plasma. For a normally incident `low intensity' femtosecond probe the absorption mechanism is predominantly inverse bremsstrahlung\cite{sandhu2005,AmitavaPOP2014}, which depends on the plasma scale-length ($L$), the electron-density ($n_{e}$) and the electron-ion collision frequency ($\nu_{ei}$)\cite{Kruerbook}. The last parameter is given by
\begin{equation}
\nu_{ei}(n_{e},T_\textrm{eV})\approx 3\times10^{-6}\ln \Lambda \frac{n_{e} Z}{T^{3/2}_\textrm{eV}}\, \textrm{s}^{-1},
\end{equation}
where $Z$ is the degree of ionization, $\ln\Lambda$ is the Coulomb logarithm, $n_{e}$ is the electron-density and $T_\textrm{eV}$ is the plasma electron temperature in units of eV. 
Given the exponential density profile  $n_e=n_{cr} \exp(-x/L)$, the electrical permittivity at a laser-frequency $\omega$ is given by 
\begin{equation}
\epsilon (x,L,\nu_{ei})=1-\frac{\exp(-x/L)}{1+i(\nu_{ei}/\omega)}.
\end{equation}
Therefore the reflectivity of the probe can be written as \cite{Kruerbook}
\begin{equation}
R \propto \exp\left(-\frac{8\nu_{ei}^{*}L}{3c}\right),
\end{equation}
where $\nu_{ei}^{*}$ is calculated at the critical-density of the probe and $L$ can be related to plasma expansion speed as $L = V_{exp} t$ and hence,
\begin{equation}
R \propto \exp\left(-t/\tau\right)
\end{equation}
where, $\tau = 3c/8\nu_{ei}^{*}V_{exp}$.
Therefore, measuring the reflectivity decay time from the experiment, it is possible to measure the $\nu_{ei}^{*}$ by knowing $V_{exp}$ or vice versa. Figure \ref{rflspike} shows the result of pump-probe reflectivity measurement (reflected probe intensity as a function of probe delay) using a BK-7 target being irradiated by intense femtosecond pump pulse ($I_{L}$ = 1.8$\times$ 10$^{17}$ W/cm$^2$, 30 fs, 800 nm). At negative probe delays, the probe reflects from the unexcited target dielectric and hence shows a low reflectivity level. Whereas on excitation by the pump pulse, the probe reflectivity spikes up to high values and then at later time it again decays down (with decay time $\tau=1.7\pm0.2$ ps) as the probe gets absorbed in the expanding plasma. An simple estimate of $\nu_{ei}^*\sim10^{14}$ s$^{-1}$  gives rise a plasma expansion speed of $V_{exp}\sim10^{7}$ cm/s. This technique helped us to explore the electron-ion collision rates at various depths of a solid density plasma gradient by simultaneously probing with 800 nm, 400 nm and 266 nm beams\cite{PrashantOPX2014}.

However, all those earlier measurements described above does not have enough temporal resolution to detect the early-time ultrafast plasma dynamics unlike shown in our recent experiment\cite{AmitavaPRL2015} where we have coupled the pump-probe Doppler spectrometry with the reflectivity measurements. At relatively higher intensity (of $9\times10^{16}$ W/cm$^2$) than that used in Adak \textit{et al.}\cite{AmitavaPRL2015}, a similar (slightly less rapid) ultrafast oscillation is found at THz frequency (fig. \ref{rfl-osc-2w}) with an overall exponential decay time of ($2.4\pm0.2$) ps.

Another important aspect is that the reflectivity measurement alone has complications explaining the dynamics of various physical parameters particularly at high, near relativistic intensity when a shock-like disturbance is observed to propagate towards the overdense plasma\cite{AmitavaPOP2014}. The reason behind is that the density scale length, in those scenarios, can't be modelled just by taking into account the simple adiabatic expansion of the plasma as was the case for early experiments\cite{LiuPRL1992}

 \begin{figure}[b]
\includegraphics[width=\columnwidth]{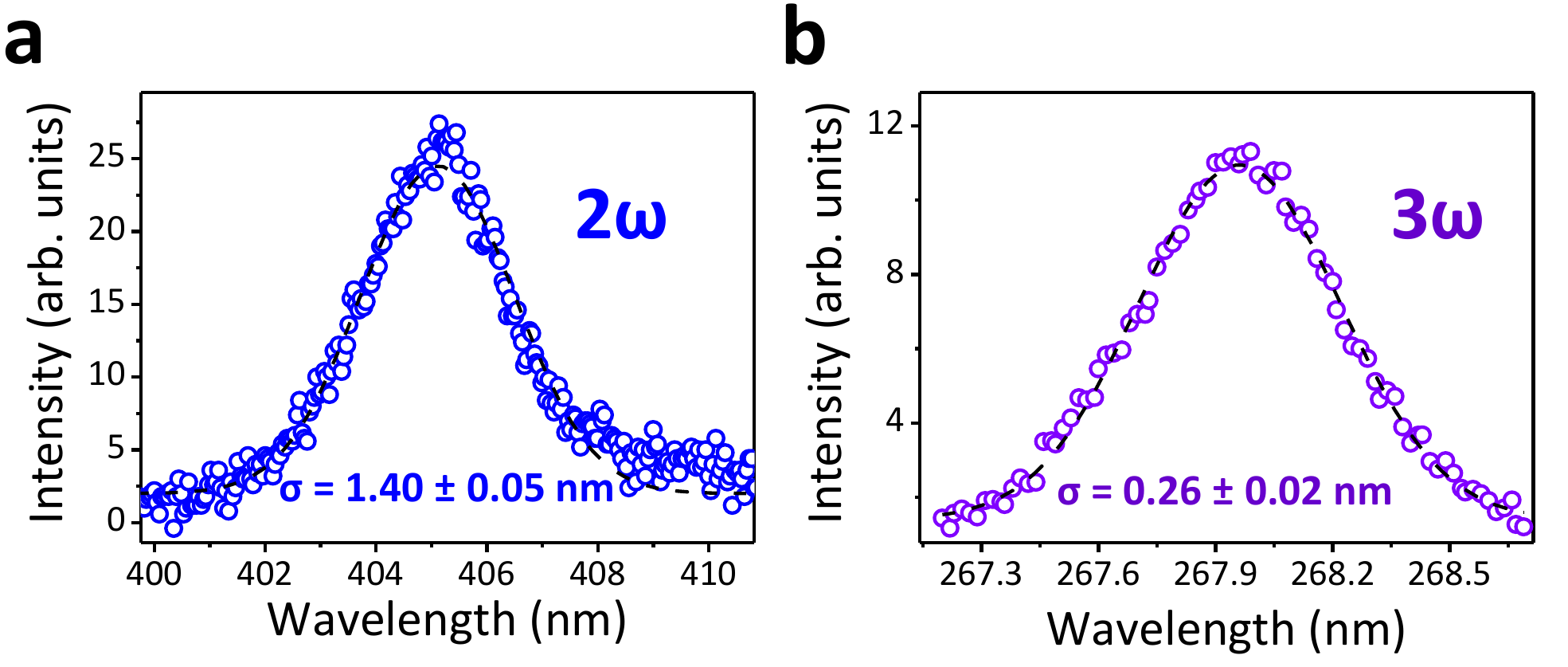}
\caption{The typical probe spectra for (a) second harmonic and (b) third harmonic probe. The black dotted lines in both cases are the respective gaussian fits.}\label{probe-spectra}
\end{figure}

\section{Pump-probe Doppler spectrometry} 

The technique of pump-probe Doppler spectrometry\cite{mondalprl2010,AmitavaPOP2014,AmitavaPRL2015} relies on the principle of the reflection of a light (probe) from a moving mirror (critical density layer or plasma mirror). Following Gjurchinovski \textit{et al.}\cite{Gjurchinovski2013}, the shifted frequency of the reflected light, incident at an angle of incidence $\alpha$ on a moving (non relativistic) mirror can be written as

\begin{equation}
\omega' = \omega_0 \frac{1+2v \cos \alpha /c + v^2/c^2}{1-v^2/c^2},
\end{equation}
where $\omega_0, v$ and $c$ are the incident light frequency, speed of the mirror moving towards the source of light and speed of light in free space respectively. In our case of near normally incident ($\cos \alpha\sim 1$) probe and small $(v/c)$, the corresponding wavelength shift ($\Delta\lambda$) can be approximated as,
\begin{equation}
\Delta\lambda=-2v/c\lambda_0
\end{equation}
Therefore, $v = -0.5 c \Delta\lambda/\lambda_0$.

In order to measure the Doppler shift $\Delta\lambda$, the central wavelength $\lambda_0$ (is found by a Gaussian fitting) of a reference probe-spectrum is subtracted from the central wavelength of the reflected probe-spectrum from the plasma. The spectrum of the fundamental laser is very broad (more than 30 nm) and is difficult to fit by a single Gaussian curve\cite{mondalprl2010}. Hence it is generally not used as probe, rather a second or third harmonic conversion is done for the probe beam. The higher harmonics can also penetrate deep inside plasma to capture the dynamics right there which would have not been possible by the fundamental wavelength. The typical second and third harmonic probe spectra used for pump-probe Doppler spectrometry experiments are shown in Fig. \ref{probe-spectra}, where the $\sigma$ values (square root of the variance) of the gaussian fit are 1.4 nm and 0.26 nm respectively. High resolution spectrometers, e.g. SpectraSuite-HR2048 for second harmonic and Avaspec-3648-USB2 for third harmonic are typically used to record the spectra at various probe delays\cite{mondalprl2010,AmitavaPOP2014}. Our earlier study\cite{AmitavaPOP2014} using this pump-probe Doppler spectrometry demonstrates that at relativistic laser intensity a shock-wave-like disturbance gets launched by the pump interaction and propagates into the target. Furthermore, the above technique using ultraviolet probe beam helped us to find out a simple method to control the shock speed just by tuning the prepulse intensity contrast of the pump laser beam\cite{AmitavaPOP2017}.

\begin{figure}[t]
\includegraphics[width=\columnwidth]{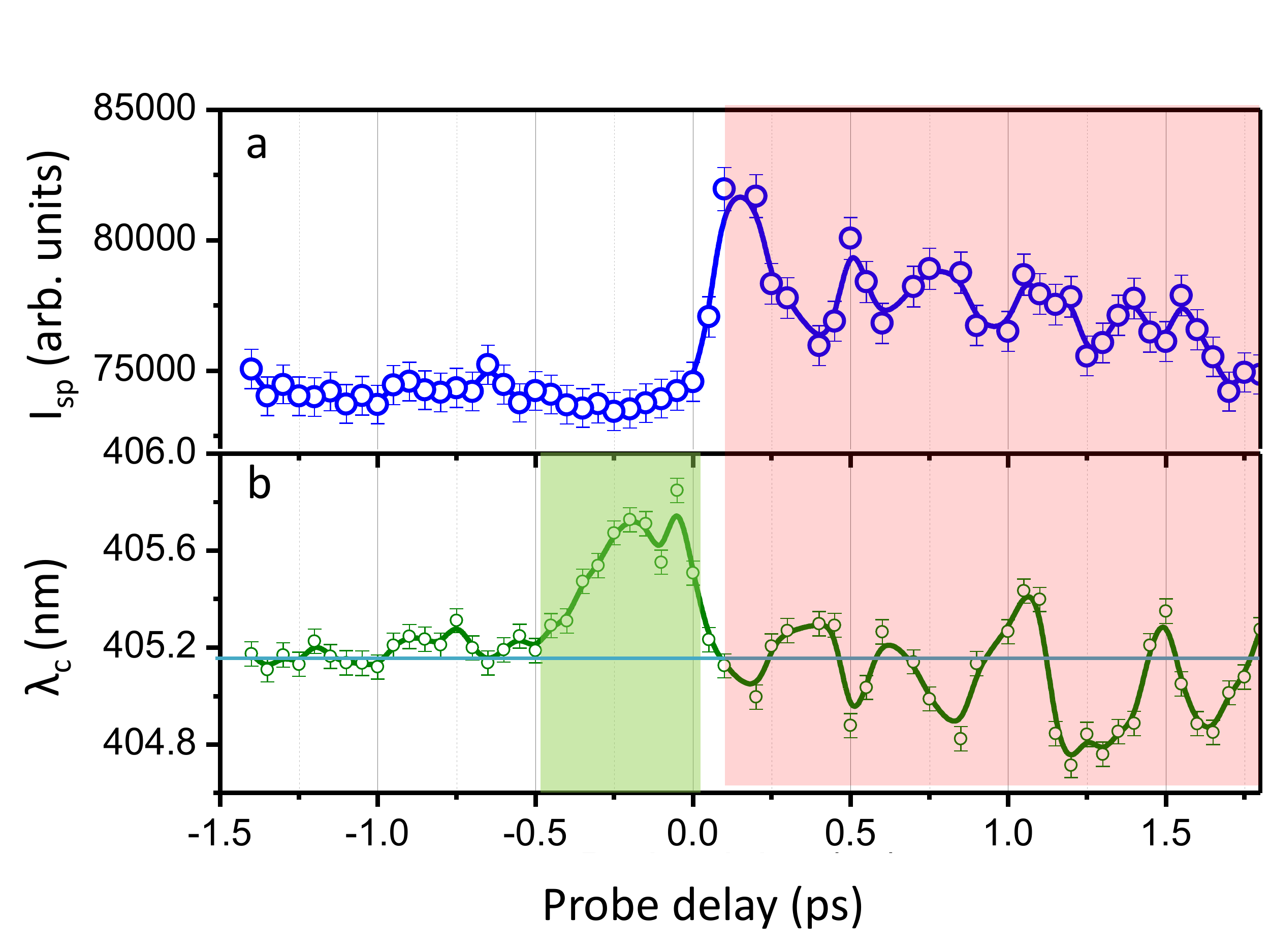}
\caption{Time resolved measurement of (a) spectral intensity (proportional to reflectivity) and (b) Doppler shift of the second harmonic probe simultaneously at a pump pulse intensity of $\sim 5\times10^{16}$ W/cm$^2$. A clear one-to-one anticorrelation between the curves in (a) and (b) is observed. The early-time red shift at negative delay could be because of the prepulse-driven density modification or a inward moving compression wave.}\label{simul-osc}
\end{figure}

In the last part of this section, we show how the simultaneous measurement of time resolved reflectivity (proportional of spectral intensity) and Doppler shift helps to study the ultrafast laser-plasma dynamics at non-relativistic laser intensity.
As was indicated in earlier section, the ultrafast plasma dynamics can be thoroughly investigated by simultaneous measurement of time resolved intensity and spectrum of the reflected probe. For this experiment the focused intensity of the excitation pump pulse on a BK-7 plane target was $0.5\times10^{17}$ W/cm$^2$. We have used a second harmonic probe (the spectrum is shown in Fig \ref{probe-spectra}a) and measured the spectral intensity (by integrating) and the central wavelength (by gaussian fit) at various probe delays. The results are shown in Fig. \ref{simul-osc}. At negative probe delays the spectral intensity (proportional to reflectivity) remains lower and flatter. Whereas, upon excitation by the pump pulse the reflected probe intensity first increases and then decreases with periodic oscillations. On the other hand, the simultaneous measurement of the central wavelength of the reflected probe spectrum shows red Doppler shit during 0.5 ps at negative delays (green shade) and then oscillates with a proper anticorrelation with the reflected probe intensity (red shade). This red shift could be because of a  picosecond-prepulse-driven plasma density modification or a inward moving compression wave. The simultaneous oscillations at positive delays are due to a generation of THz acoustics because of a ultrafast hydrodynamic evolution of the hot, dense plasma as discovered in our earlier study\cite{AmitavaPRL2015}. The back and forth oscillation of the probe critical surface riding on the ultrafast acoustic oscillation is the cause of the periodic red and blue Doppler shift.
\section{Discussion}
Time resolved pump-probe reflectometry and spectrometry were discussed in detail. These techniques helped to find out ultrafast dynamics in solid density plasmas on sub 100 fs time scales to several ps time scales. We have also explained how they are useful in order to find out ultrafast optics at various depth of the plasma gradient, exploring shock-like disturbance in solid dense plasmas and how to control the shock speed in it. The simultaneous measurement of reflectometry and spectrometry helps us to investigate a new THz acoustics phenomena due rapid hydrodynamics of the laser-driven dense plasma. 

\section{Acknowledgements}
GRK acknowledges a  J.C. Bose Fellowship grant.

\end{document}